# Desensitization of a balance with Langmuir binding of weights

Richard Lanzara
*Bio Balance, Inc. 30 West 86th Street New York, N.Y. USA 10024-3600*

A balance is described with Langmuir binding of weights to each side. This simple physical system, when coupled with a previously derived equation for the equivalent displacement of weight that perturbs the equilibrium of a balance, models two-state, receptor systems in pharmacology for receptor activation. Surprisingly, this system also shows desensitization similar to many drug-receptor systems. Although understanding the biophysics behind receptor signaling is an important endeavor by itself, this analogy with a simple balance may bring new insights into other areas of physics as well.

PACS numbers: 82.53, 87.10, 82.20

Our senses function through a distinct class of molecules known as G protein-coupled receptors (GPCRs). These receptors sense our environment and control most of our physiology. They are targets for about one-third of modern drug development. The responses of many of these GPCRs to stimuli typically show hyperbolic, dose-response curves that approach a maximum value; peak and then decline back to baseline values. This phenomenon often appears as either normal or skewed bell-shaped, dose-response curves. Pharmacologists have named this decline towards baseline "desensitization" [1-15]. Although many of the most important biological events show receptor desensitization, this phenomenon remains poorly understood [1-12]. Desensitization also appears counterintuitive by suggesting that more of an activating drug or molecule lessens the response. Therefore, it remains difficult for pharmacologists to reconcile how a molecule can both activate and deactivate those same receptors, often within a millisecond to microsecond time frame [12]. If there were a simple physical system that shows a similar response, it may offer important insights into this problem.

A simple balance may offer such a physical system. The balance can be conceptualized to represent receptor equilibrium and has the additional advantage of revealing relationships between a physical and chemical, two-state equilibrium [14-16]. When a receptor chemically "senses" a hormone or drug molecule, the molecule binds preferentially to one state of the receptor (or one side of the balance) and perturbs the initial receptor equilibrium (so that it is out of balance) [14,15]. This perturbation produces an induced shift in the initial receptor equilibrium toward the preferred binding state. This has the effect of enriching the population of the preferred receptor state. For a cellular signal to be transmitted, there must also exist a series of poised equilibria that transmit these perturbations from the external signals into intracellular messages.

The quantifiable nature of this perturbation or shift is analogous to a previously derived equation for a weight, $\Delta w$, that equivalently perturbs a balance in equilibrium,

$$\Delta w = \frac{S_1 w_2 - S_2 w_1}{w_1 + S_1 + w_2 + S_2} \tag{1}$$

where $S_1$ and $S_2$ are the supplemental weights added to each side of a simple balance and $w_1$ and $w_2$ are the weights on the balance in an initial equilibrium before the addition of the supplemental weights [13]. The nature of $\Delta w$ provides a mechanism to calculate perturbations in similar two-state, equilibrium systems [13-15].

By introducing Langmuir binding equations for $S_1$ and $S_2$ into Equation (1), we can transform a balance into a biophysical receptor model that describes many aspects of drug-receptor interactions [13-16]. Letting $w_1 = R_1$, $w_2 = R_2$ where $R_1$ and $R_2$ represent the maximum amount of weight allowable for each side. Substituting the Langmuir binding expressions for the supplemental weights, $S_1=R_1(S)/(S+K_1)$, and $S_2=R_2(S)/(S+K_2)$ into equation (1), where $K_1$ and $K_2$ are the affinity constants for the two sides of the balance, and letting $\Delta w = \Delta R$ yields,

$$\Delta R = \frac{R_1 R_2 (S)(K_2 - K_1)}{R_1(2S+K_1)(S+K_2) + R_2(S+K_1)(2S+K_2)} \qquad (2)$$

where $\Delta R$ represents the change in the amount of weight equivalent to the perturbation of the initial equilibrium by unequal ($K_1 \neq K_2$) molecular binding. This equation is identical to equation (6) in reference [14] that was originally derived for drug-receptor responses where "$\Delta RH$" represented the change in the initial receptor states.

This model differs from other two-state receptor models in that the other models do not calculate the net change in the amount of the receptor that is shifted by the preferred binding. This is primarily because the other models do not make the net change in the shift an unknown variable that can be solved explicitly. This model corrects this problem by solving explicitly for the net shift, $\Delta R$ [13-15].

Previously, Equation (2) was used to model dose-response relationships in pharmacology [14,15]. However before this study, Equation (2) was not subjected to an experimental test with a physical balance. For the sake of inquiry, Equation (2) was tested with a simple two-pan balance. $K_1$ and $K_2$ were set equal to 10 g and 100 g respectively. They represent the unequal affinities ($1/K_1$ and $1/K_2$) of each pan (A and B) for (S). $R_1$

and $R_2$ were each set arbitrarily equal to 100 g, and the amount of (S) was allowed to vary up to 500 for this particular experiment. Note that for this demonstration (S) represents an amount of weight (or a concentration) potentially available for binding up to the saturating conditions of $R_1$ and $R_2$. It should be noted that the Langmuir binding equations for $S_1$ and $S_2$ allow a fraction of the available maximum weights $R_1$ and $R_2$ to weight pans A and B respectively (see Table 1). The Langmuir binding equations are plotted for $S_1$ and $S_2$ in Figure 1. As seen in Figure 1 and Table 1, the theory fits the experimental data for the balance very well.

The insert in the upper right corner of Figure 1 is the theory plotted on a reverse axis with an expanded scale. From the physical point of view, the insert demonstrates the striking similarity to similar figures from the tracings of excitatory postsynaptic currents produced in neurons by the neurotransmitter, molecule glutamate [12], and similar pharmacological dose-response curves [4,10,14] (also see below). This demonstrates the similarity of response curves produced with this balance model to the nonlinear response curves that are ubiquitous in many complex cellular systems. Although this is an arbitrary physical system, it could easily be made more realistic by decreasing the constants $K_1$ and $K_2$ to create a more realistic system for the binding of molecules to receptors on a cell surface [14, 15]. From a purely physical perspective, these observations suggest that suitable physical systems may be created to mimic the behaviors of the more complex biological ones.

As a further example, Figure 2 shows the balance as a slightly more complicated but more realistic model for a hypothetical cellular receptor. The parameter, (S), represents the concentration of molecules in solution and $\Delta R$ is the theoretical response calculated by Equation (2). Also plotted in Figure 2 are the Langmuir binding plots for $S_1$ and $S_2$ and the total binding on a logarithmic scale. Desensitization occurs in the presence of continued binding and is observed as a bell-shaped curve as seen for ΔR in Figure 2.

Another interesting observation is that the response of the system occurs at only a fraction of the total binding (compare 50% ΔR response with the curve for S1+S2 in FIG. 2). This has been called the phenomenon of "spare receptors" in pharmacology and has puzzled pharmacologists for many years. However, we can now see what produces the apparent spare receptor reserve. Since only a fraction of the receptor molecules (or weight in the case of the balance) is transferred as an equivalent net shift or response, ΔR, the magnitude of this response will always be some smaller fraction of the total receptor pool or weight of the system. Interestingly, this also demonstrates that a balance can show the curiosity, which is known as "spare receptors", when modeled with Langmuir binding.

The peak response of these curves can be found by taking the first derivative of Equation (2) with respect to S and setting it equal to zero. The peak occurs at $(K_1 K_2/2)^{1/2}$ which can be useful to know for modeling and experimental purposes. Note that inhibitors can also be introduced into Equation (2) by including the inhibition factor $(1+ I/K_i)$ for an antagonist (I with its dissociation constant, $K_i$) multiplied times each of the dissociation constants, $K_1$ and $K_2$. This reproduces the effects of inhibitors in pharmacological systems and leads to the discovery of many useful effects [14].

Since the manner by which a biological receptor functions can be modeled by this approach [13-15], pharmacologists and physiologists may find Figure 3 interesting for its similarity to the electrophysiological measurements of excitable cells [2,4,10,11,12]. The family of curves displayed in Figure 3 demonstrate the effect of decreasing the available receptor pool ($R_1$ and $R_2$) by 20 to 80%. The similarity between the responses observed in complex pharmacological systems and the balance model strongly suggests that cellular receptors mimic miniature chemical balances in a coupled, two-state equilibrium that can be shifted by perturbations from unequal molecular binding, or possibly other forces. It also suggests that many of the complex behaviors observed in these biological responses and similar systems are a direct result of the physicochemical responses to the perturbations affecting the underlying equilibrium of these systems. Although not explored here, this may also suggest potential applications in the areas of rapid chemical kinetics and theory of enzyme reactions.

In conclusion, treating a simple balance with the restriction of Langmuir binding creates systems that behave surprisingly similar to many receptor systems that desensitize and suggests that a simple physical model displays those characteristics previously thought peculiar to many complex biological processes [1-15]. Desensitization is found in many unusual and amazing places - our senses, drug receptors, the neurochemical synapses within our brains, and a simple balance. This may be the first time that the behavior of a balance was described as a desensitizing system and experimentally tested. A balance that shows responses that desensitize similar to biological receptors offers a physical link between GPCR activation and desensitization, and may have implications for other areas of physics as well.

**References:**

[1] C. S. Pao and J. L. Benovic, Science's STKE, http://www.stke.org/cgi/content/full/sigtrans; 2002/153/pe42 (2002).

[2] M. T. Bianchi, K. F. Haas, and R. L. Macdonald, J. Neurosci. **21(4)**, 1127–1136 (2001).

[3] N. R. Sullivan Hanley and J. G. Hensler, JPET **300**, 468–477 (2002).

[4] K. M. Partin, J. Neurosci. **21(6)**, 1939–1948 (2001).

[5] M. He, *et al.*, Mol Pharm. **62**, 1187–1197 (2002).

[6] B. January, *et al.*, JBC **272**, 23871–23879 (1997).

[7] K. Bender, *et al.*, JBC **276**, 28873-28880 (2001).

[8] C. Blanchet and C. Lüscher, PNAS **99**, 4674–4679 (2002).

[9] M. Bünemann, *et al.*, Annu. Rev. Physiol. **61**, 169–92 (1999).

[10] J. S. Marchant and C. W. Taylor, Biochem. **37**, 11524-11533 (1998).

[11] M.V. Jones and G.L. Westbrook, Trends in Neurosci. **19**, 96-112 (1996).

[12] B. Sakmann, Neuron **8**, 613-629 (1992).

[13] R. Lanzara, Math. Biosci. **122**, 89-94 (1994).

[14] R. Lanzara, *A Method for determining drug compositions to prevent desensitization of cellular receptors*. United States Patent **5,597,699** (Jan. 28, 1997).

[15] R. Lanzara, Can. J. Physiol. & Pharm. **72**, 559a (1994).

[16] L. Rubenstein and R. Lanzara, J Mol. Struct. (Theochem) **430/1-3**, 57-71 (1998).

**Table 1. Experimental and theoretical values for the balance.**

| (S)* | Pan A weight (g)* | Pan B weight (g)* | measured shift (deg.) | shift (% normalized) | theory ($\Delta R$) (% normalized) |
|---|---|---|---|---|---|
| 1 | 9 | 1 | 18 | 37 | 40 |
| 10 | 50 | 9 | 43 | 88 | 91 |
| 30 | 75 | 23 | 49 | 100 | 100 |
| 50 | 83 | 33 | 47 | 96 | 95 |
| 100 | 91 | 50 | 39 | 80 | 82 |
| 300 | 97 | 75 | 27 | 55 | 56 |

* Please note that (S) represents a "concentration of available weight" that "weights" each of the pans A and B according to the Langmuir equations for $S_1$ and $S_2$. The maximum weight available for each pan is given by $R_1$ and $R_2$ respectively.

FIG. 1. Plots of the Langmuir binding to pans A and B of the balance with the theory (Theory ΔR) from Eq. (2) compared with the experimental measurements (experiment shift). The insert is the same curve as the theory (ΔR) plotted with a reverse y-axis and an expanded scale.

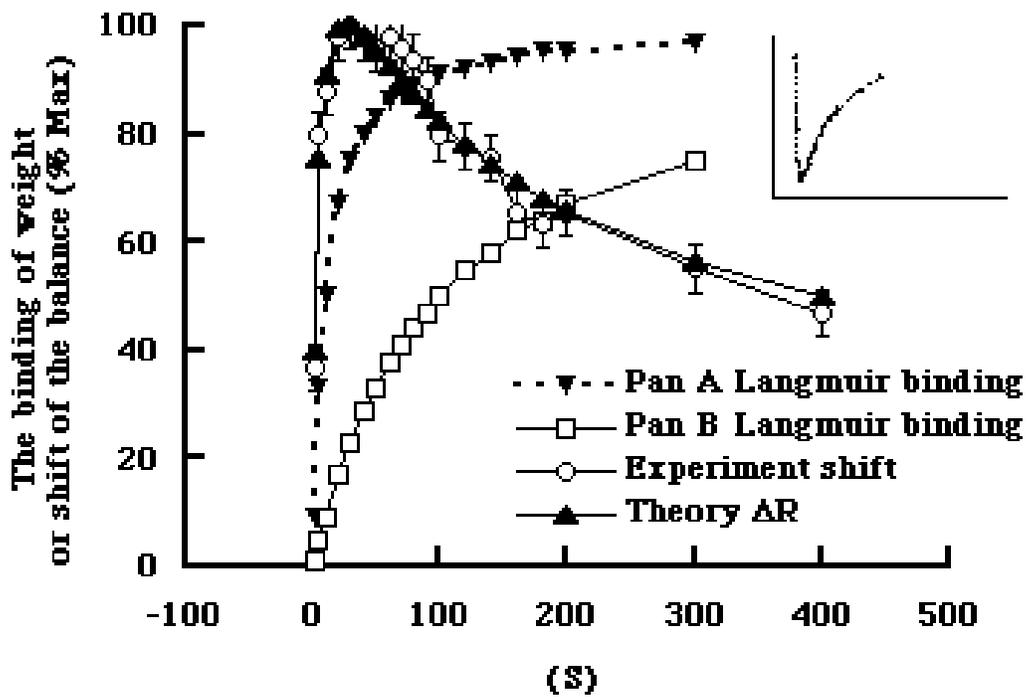

FIG. 2. Shows Langmuir binding plots of $S_1$ and $S_2$ for $K_1$ and $K_2$ of $1 \times 10^{-9}$ and $1 \times 10^{-7}$ respectively. The parameter, (S), represents the concentration of molecules in solution. The total binding is given as S1+S2, and $\Delta R$ is the theoretical response from Eq. (2). Note that at 50% of the response ($\Delta R$) the total binding is only 20% or less.

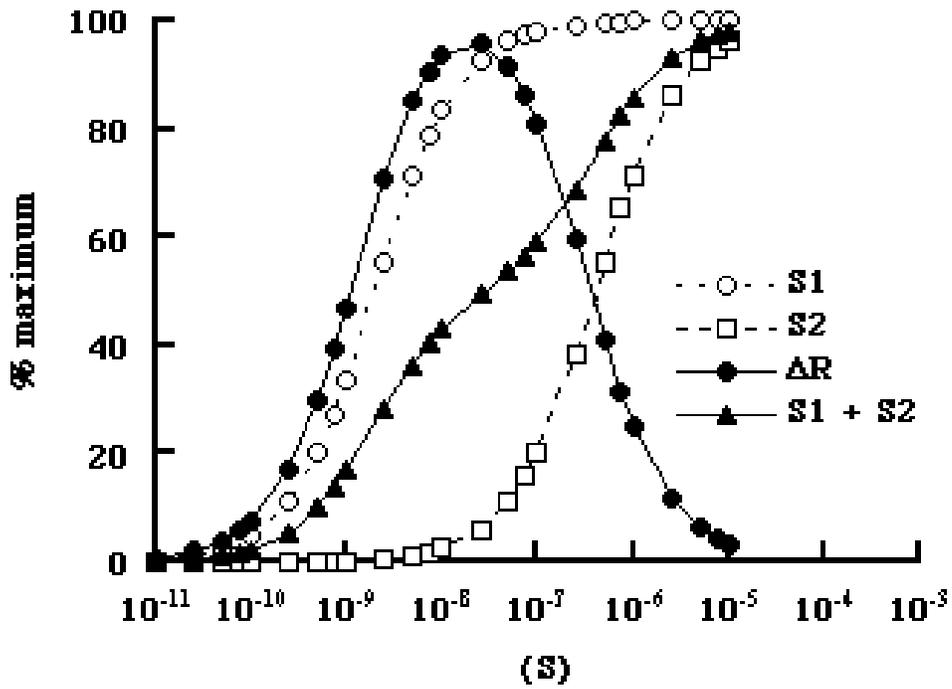

FIG. 3. The plot for a family of curves that demonstrate the response curve (ΔR) from FIG. 2 with 80%, 50%, and 20% of the total receptor pool ($R_1$ and $R_2$) on a reverse axis with an expanded scale. The insert shows ΔR plotted at a larger scale.

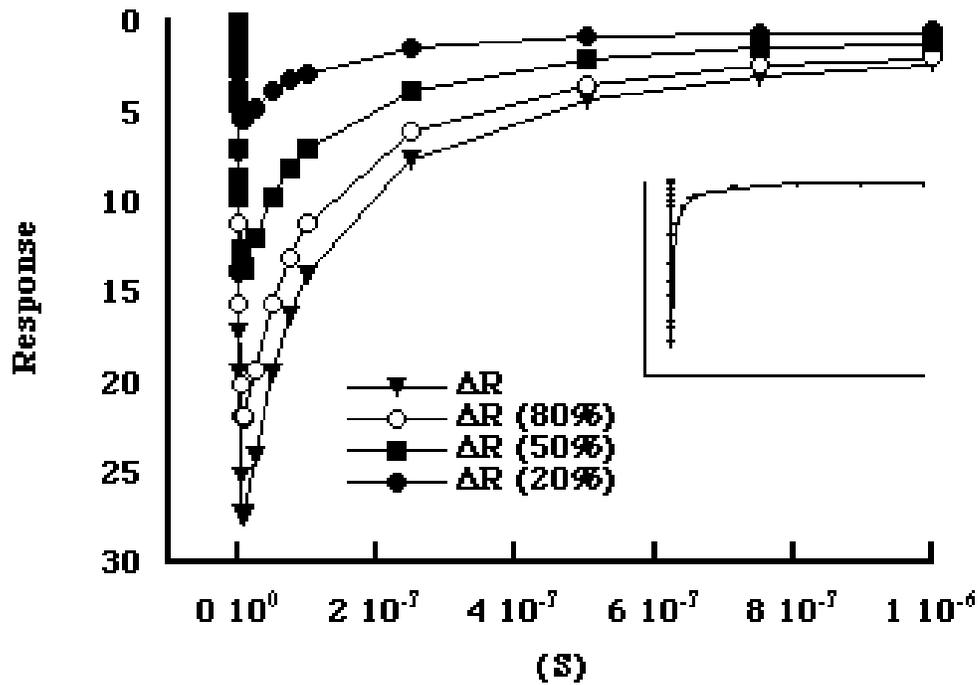